\documentclass[preprint,1p,twocolumn]{elsarticle}

\journal{Physica A}

\begin{document}
\begin{frontmatter}

\title{Fermi--Pasta--Ulam--Tsingou problems: Passage from Boltzmann to $q$-statistics}

\author[add1]{Debarshee Bagchi}
\author[add1,add2,add3]{Constantino Tsallis}
\ead{tsallis@cbpf.br}

\address[add1]{Centro Brasileiro de Pesquisas Fisicas - Rua Xavier Sigaud 150, 22290-180 Rio de Janeiro-RJ,  Brazil}
\address[add2]{National Institute of Science and Technology of Complex Systems - Rua Xavier Sigaud 150, 22290-180 Rio de Janeiro-RJ,  Brazil}
\address[add3]{Santa Fe Institute - 1399 Hyde Park Road, New Mexico 87501, USA}
\address[add4]{Complexity Science Hub Vienna, Josefst\"adter Strasse 39, 1080 Vienna, Austria}

\begin{abstract}
The Fermi-Pasta-Ulam (FPU) one-dimensional Hamiltonian includes a quartic term which guarantees 
ergodicity of the system in the thermodynamic limit. Consistently, the Boltzmann factor
$P(\epsilon) \sim e^{-\beta \epsilon}$ describes its equilibrium distribution of one-body
energies, and its velocity distribution is Maxwellian, i.e., $P(v) \sim e^{- \beta v^2/2}$. 
We consider here a generalized system where the quartic coupling constant between sites decays 
as $1/d_{ij}^{\alpha}$ $(\alpha \ge 0; d_{ij} = 1,2,\dots)$. Through {\it first-principle} 
molecular dynamics we demonstrate that, for large $\alpha$ (above $\alpha \simeq 1$), i.e., 
short-range interactions, Boltzmann statistics (based on the {\it additive} entropic functional 
$S_B[P(z)]=-k \int dz P(z) \ln P(z)$) is verified. However, for small values of $\alpha$ (below
$\alpha \simeq 1$), i.e., long-range interactions, Boltzmann statistics dramatically fails and 
is replaced by q-statistics (based on the {\it nonadditive} entropic functional $S_q[P(z)]=k 
(1-\int dz [P(z)]^q)/(q-1)$, with $S_1 = S_B$). Indeed, the one-body energy distribution is 
q-exponential, 
$P(\epsilon) \sim e_{q_{\epsilon}}^{-\beta_{\epsilon} \epsilon} \equiv [1+(q_{\epsilon} - 1)
\beta_{\epsilon}{\epsilon}]^{-1/(q_{\epsilon}-1)}$ with $q_{\epsilon} > 1$, and its velocity 
distribution is given by $P(v) \sim e_{q_v}^{ - \beta_v v^2/2}$ with $q_v > 1$. Moreover, within 
small error bars, we verify $q_{\epsilon} = q_v = q$, which decreases from an extrapolated value 
q $\simeq$ 5/3 to q=1 when $\alpha$ increases from zero to $\alpha \simeq 1$, and remains q = 1 thereafter.
\end{abstract}


\end{frontmatter}

%% main text
Ludwig Boltzmann intensively tried to derive his celebrated weight, generalized by Gibbs into what 
is currently called the Boltzmann-Gibbs (BG) weight, from Newtonian mechanics and no other hypothesis. 
This is sometimes referred to as the {\it Boltzmann program}. The nonlinearity of the entangled particle
dynamics makes the task a formidable one, and Boltzmann did not achieve it. Even today this remains as a
very basic unsolved mathematical problem.  This in no way means that we do not have a quite neat scenario 
about the validity of the Boltzmann weight. Indeed, it is clear by now that if the Newtonian dynamics of 
the system is such that its maximal Lyapunov exponent is positive ({\it strong chaos}), then the dynamics 
is mixing, hence ergodic, and it is on this basis (along the lines of the so-called {\it Stosszahlansatz, 
molecular chaos hypothesis}) that  BG statistical mechanics is constructed. The situation is much more 
complex when the  maximal Lyapunov exponent virtually vanishes ({\it weak chaos}). This is the discussion 
that we address here through a paradigmatic system, namely the Fermi-Pasta-Ulam (FPU) Hamiltonian \cite{FPU1955} 
(see details in \cite{ZabuskyKruskal1965,Gallavotti2008,OnoratoVozellaPromentLvov2015}). This system plays a 
relevant role in the discussion of Fourier's law for heat flow (heat conductivity), rapid (or slow) sharing 
of energy, eventually yielding equipartition of energy and the zeroth law of thermodynamics. It consists of 
a  ring (chain with periodic boundary conditions) of oscillators which, in addition to their kinetic energies, 
interact through both harmonic and anharmonic terms. We focus on the following Hamiltonian
\cite{ChristodoulidiTsallisBountis2014,BagchiTsallis2016,ChristodoulidiBountisTsallisDrossos2016}:
\begin{equation}
\mathcal{H} = \sum_{i=1}^N \frac {p_i\,^2}{2 m}  + \frac a {2} \sum_{i=1}^N (x_{i+1} - x_i)^2 
+ \frac {b} {4 \tilde N} \sum_{i} \sum_{j \ne i} \frac {(x_i -  x_j)^4} {d_{ij}^{~\alpha}},
\label{H}
\end{equation}
where $x_i$ and $p_i$ are the displacement and momentum of the $i$-th particle with mass $m$ (from now on, 
without loss of generality, we can set $m=1$, hence the momenta coincide with the velocities, i.e., $p_i=v_i$); 
$a \ge 0$, $b > 0$, and $\alpha \ge 0$. Here $d_{ij} = 1,2,3,...$, is the shortest distance between the $i$-th 
and $j$-th lattice  sites ($1 \le i,j \le N$). If $\alpha >1$ ($0 \le \alpha \le 1$)  we refer to short-range 
(long-range) interactions in the sense that the potential energy per particle is integrable (diverges) as 
$N \to \infty$; in particular, the limit $\alpha \to \infty$ corresponds to first-neighbor quartic interactions 
(i.e., the historical FPU $\beta$-model), and the $\alpha=0$ value corresponds to a typical mean-field scenario. 
The Hamiltonian is made (formally) extensive for all values of $\alpha$ by adopting the scaling factor 
\cite{AnteneodoTsallis1998,CirtoAssisTsallis2013,ChristodoulidiTsallisBountis2014} $\tilde N \equiv \sum_{i =1}^{N} 
\frac{1}{d_{ij}^{~\alpha}}$, which depends on $(\alpha, N)$. Note that for $\alpha = 0$ we have $\tilde N = N$, which 
recovers the rescaling usually introduced in mean-field approaches, sometimes referred to as the Kac prescription factor. 
In the thermodynamic limit $N \to\infty$, $\tilde N$ remains constant, namely $1/(\alpha-1)$, for $\alpha >1$, whereas 
$\tilde N \sim \frac{N^{1-\alpha}}{1-\alpha}$ for $0 \le \alpha <1$, and $\tilde N \sim \ln N$ for $\alpha = 1$. It can 
be verified that the introduction of $\tilde N$ in Hamiltonian (\ref{H}) is equivalent to a simple rescaling of time 
\cite{AnteneodoTsallis1998}.

We numerically integrate the Newton's equations of motion using the symplectic velocity Verlet algorithm \cite{vel-Verlet} 
with a small time-step $\Delta t \leq 10^{-2}$ such that the deviation of total energy in an isolated system is of the order
of $10^{-4}$ or less, until large times $t = 10^8$, depending on the system size $N$. The initial conditions for the displacement 
variables are randomly chosen from a uniform distribution and the momenta from a normal distribution with unit variance, both 
centered around zero. Starting from a single random initial condition, the system is evolved for $t = 10^5$ to allow the
system to fully relax to its stationary (or quasi-stationary) state before starting averaging the steady state quantities for
$t = 10^3$ time-steps (here time $t$ is measured in units of $\Delta t$).

\begin{figure}
%\centering
\includegraphics[width=5cm,angle=-90]{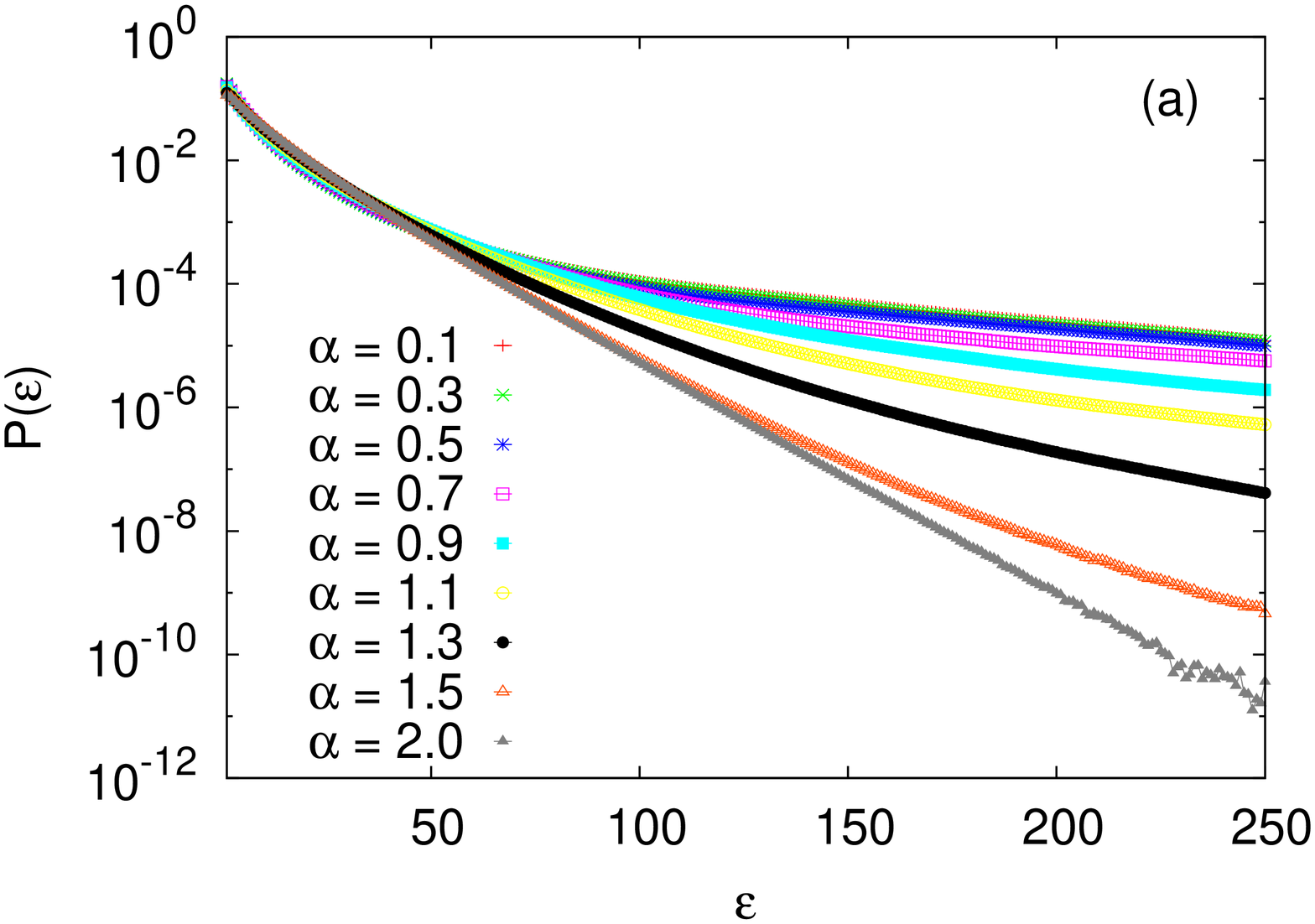}
\includegraphics[width=5cm,angle=-90]{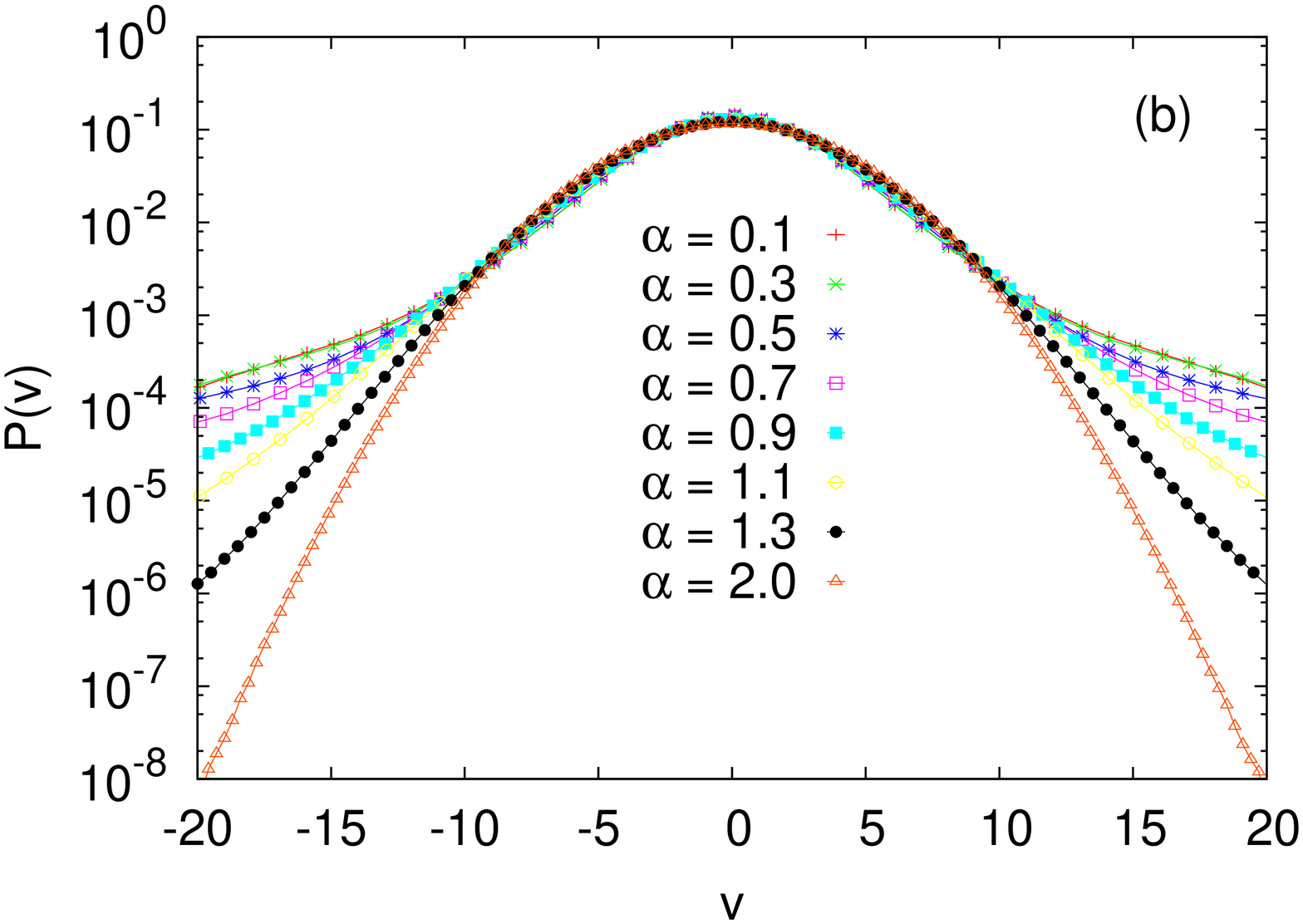}
\caption{Evolution of (a) one-particle energy $\epsilon$ distribution $P(\epsilon$) and (b) one-particle velocity $v$ 
distribution $P(v)$ in the presence of $\alpha$-ranged anharmonic interactions for different values of $\alpha$ corresponding 
to a typical case, namely $(a,b)=(1,10)$, $u=9$, and $N=8000$.
}
\label{ProbEV}
\end{figure}

First we compute the distributions of the one-particle energy 
$\epsilon_i = \frac {p_i\,^2}{2 m}  + \frac a {4} [(x_{i+1} - x_i)^2 + (x_i - x_{i-1})^2] + \frac {b} {8 \tilde N} 
\sum_{j} \frac {(x_i -  x_j)^4} {d_{ij}^{~\alpha}}$ and of velocity $v_i$ as the long-range parameter $\alpha$ is 
increased from zero on ($1 \le i\le N$). The energy and velocity distributions, $P(\epsilon)$ and $P(v)$ (for a 
homogeneous system $\epsilon_i = \epsilon$ and $v_i = v$), are obtained by collecting $\epsilon_i$ and $v_i$
of all the particles in the system and are shown in Fig. \ref{ProbEV}, a and b respectively.
\begin{figure}
%\centering
\includegraphics[width=5cm,angle=-90]{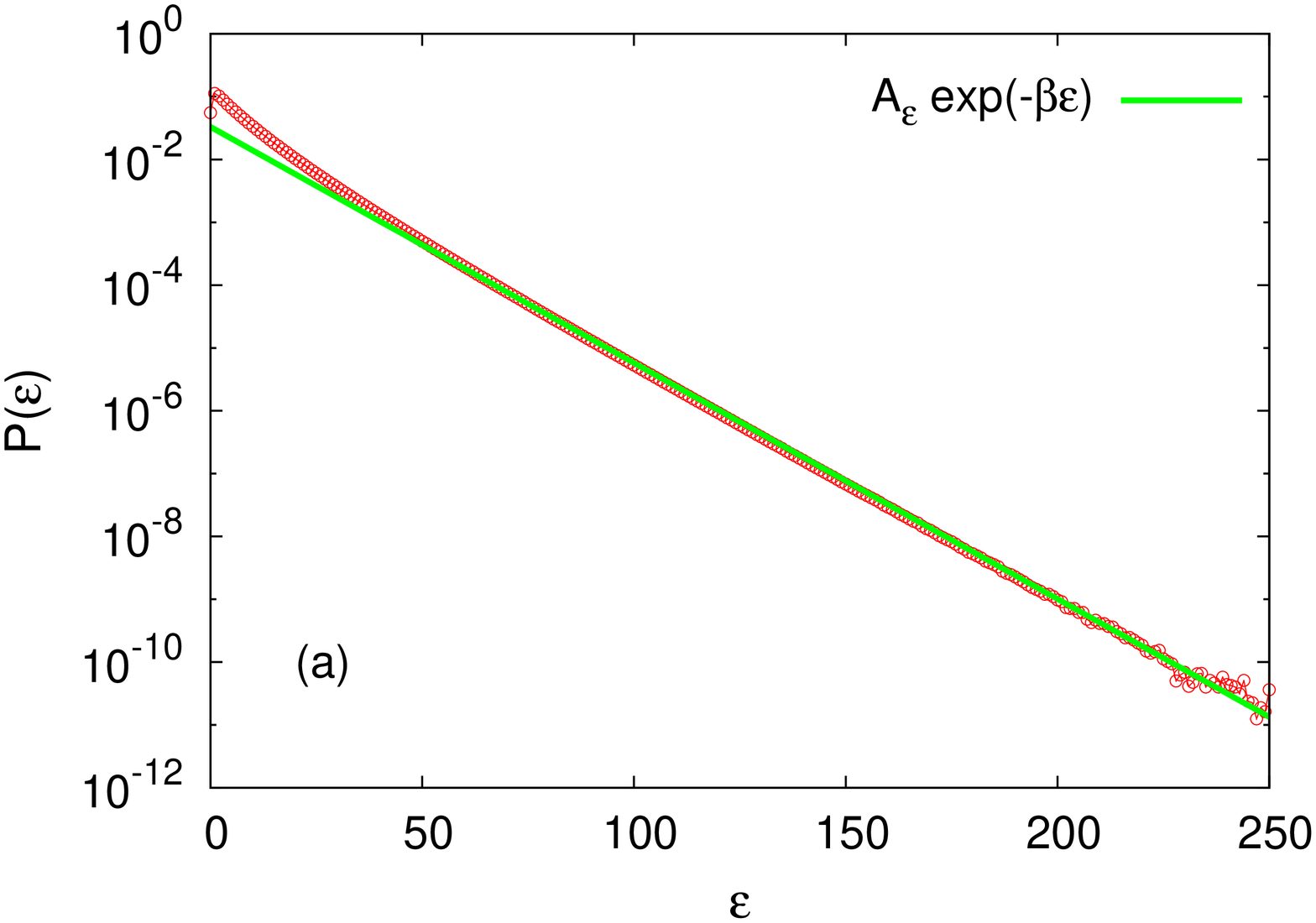}
\includegraphics[width=5cm,angle=-90]{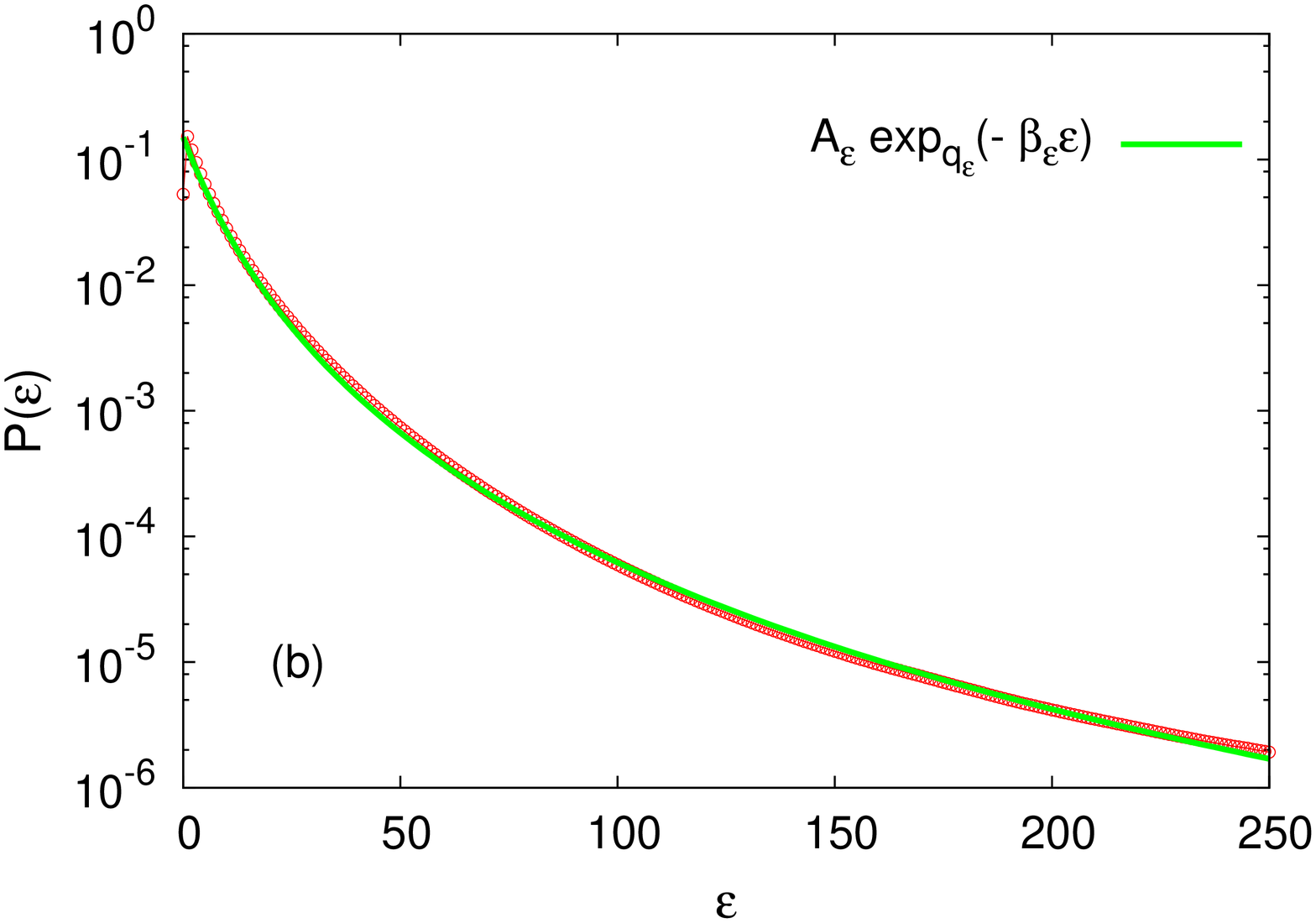} \\
\includegraphics[width=5cm,angle=-90]{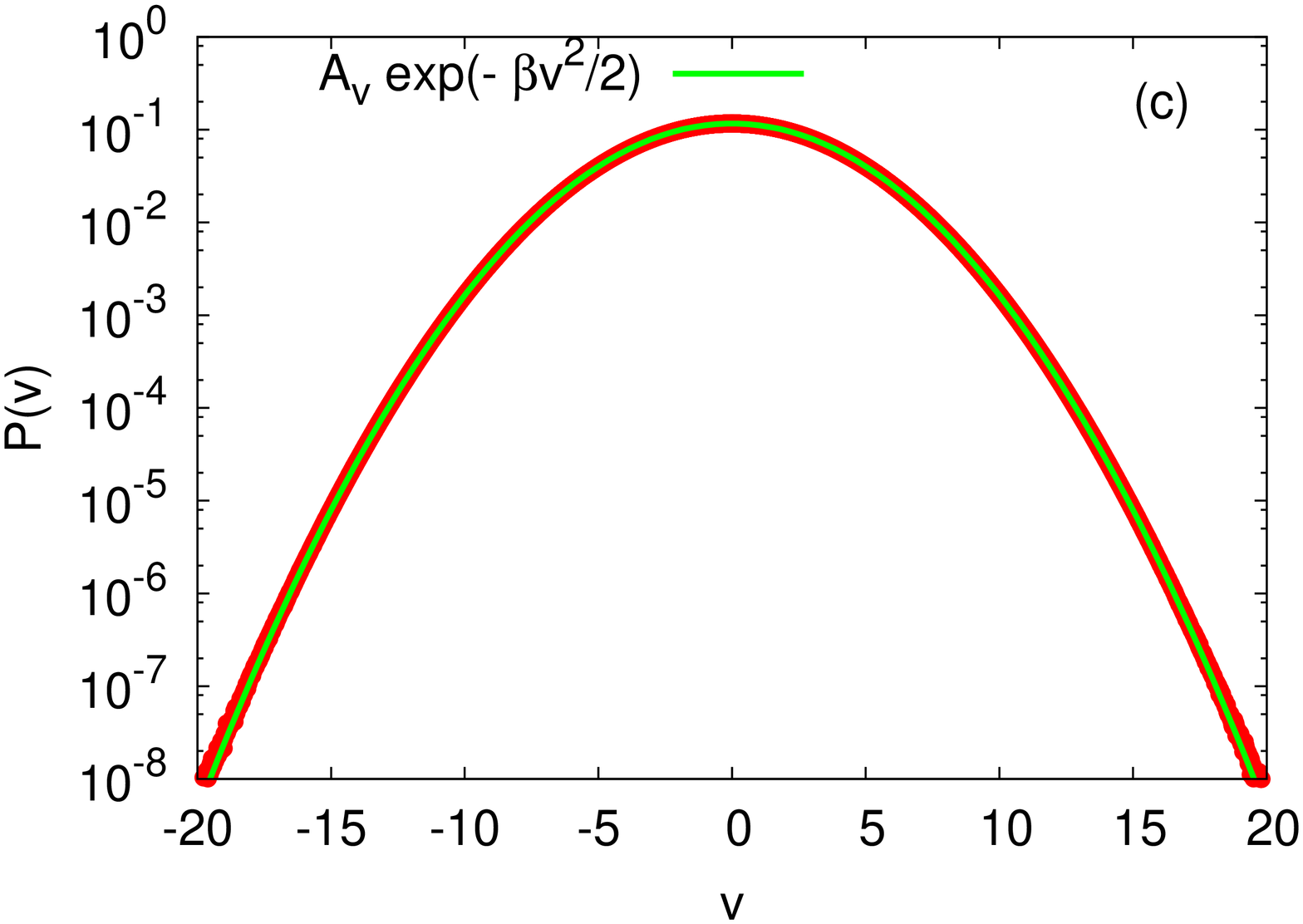}
\includegraphics[width=5cm,angle=-90]{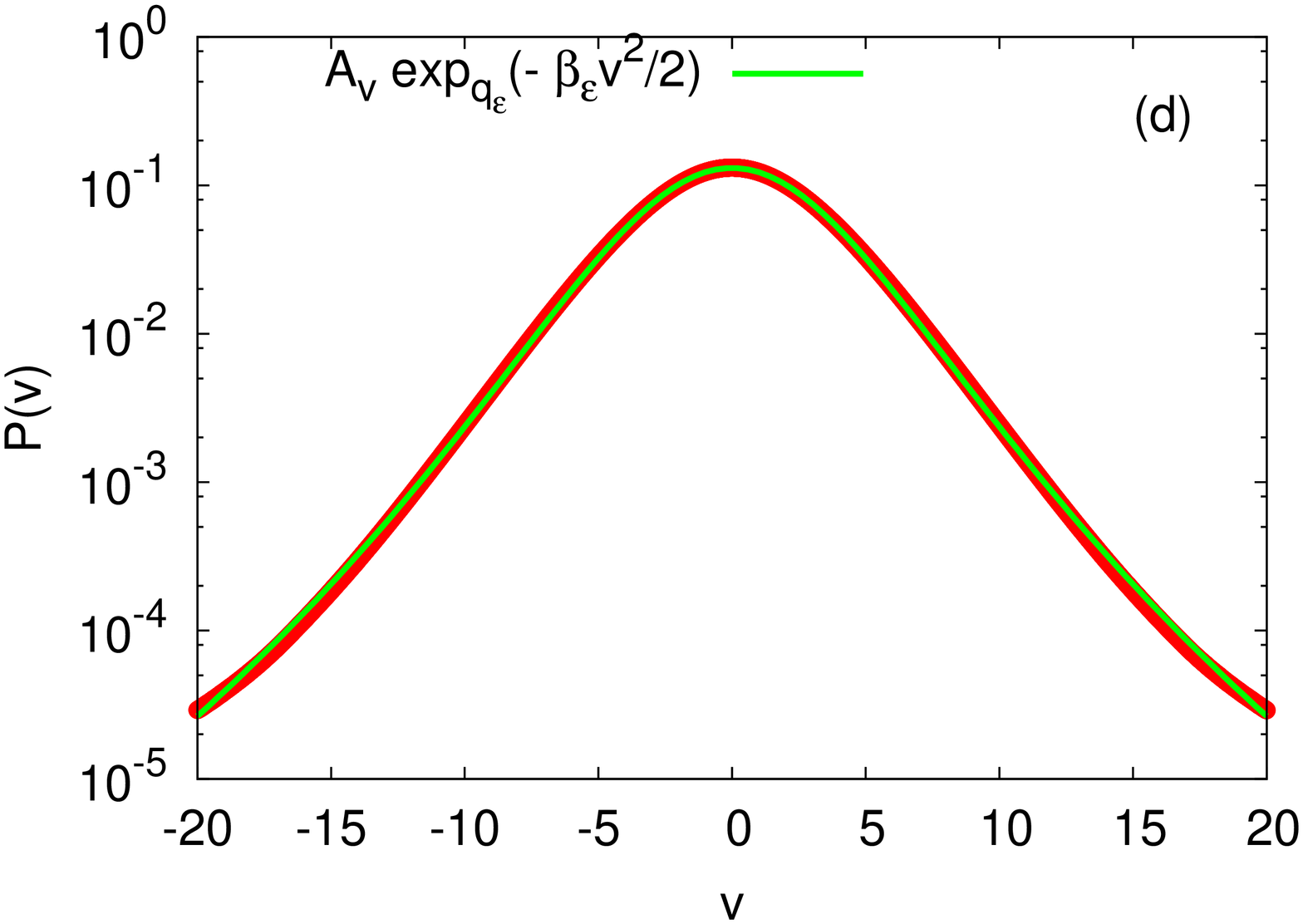}
\caption{(a) Boltzmann exponential continuous fitting of the one-particle energy $\epsilon$ distribution in the 
presence of short-range anharmonic interactions ($\alpha=2$): $P(\epsilon)=0.033\, e^{-0.086\,\epsilon}$. (b) 
The $q$-exponential continuous fitting of the one-particle energy $\epsilon$ distribution in the presence of 
long-range anharmonic interactions ($\alpha=0.9$): $P(\epsilon)=0.15\, e_{1.22}^{-0.21\,\epsilon}$. For small 
values of $\epsilon$, a slight departure from the purely exponential (or $q$-exponential) behavior is observed, 
as expected due to a regular density of states. 
(c) Gaussian continuous fitting of the one-particle velocity $v$ distribution in the presence of short-range 
anharmonic interactions ($\alpha=2$): $P(\epsilon)=0.116\, e^{-0.084\,v^2/2}$.  (d) The $q$-Gaussian continuous 
fitting of the one-particle energy $\epsilon$ distribution in the presence of long range anharmonic interactions 
($\alpha=0.9$): $P(\epsilon)=0.131\, e_{1.23}^{-0.132\, v^2/2}$.
The other parameters are the same as in Fig. \ref{ProbEV}.}
\label{qstatistics}
\end{figure}
For large $\alpha$ (short-range interactions) the energy distribution does recover the expected Boltzmann distribution. 
However, for small values of $\alpha$ (long-range interactions), the celebrated exponential distribution dramatically 
fails, and is replaced (within a fairly good numerical precision) by a $q_{\epsilon}$-exponential one, where the index 
$q_{\epsilon}$ depends on $\alpha$; see Figs. \ref{qstatistics}a and \ref{qstatistics}b where we show our simulation 
data fitted to the theoretical curves for two typical $\alpha$ values, $\alpha = 0.9, 2.0$ corresponding to the long-range 
and the short-range regimes respectively. Likewise, for the single particle velocity distribution, one obtains a Maxwell's
velocity distribution for large $\alpha$ whereas for small $\alpha$ the velocity histogram can be well approximated by a
q-Gaussian function; this is shown in Figs. \ref{qstatistics}c and d respectively.

\begin{figure}
%\centering
\includegraphics[width=5cm,angle=-90]{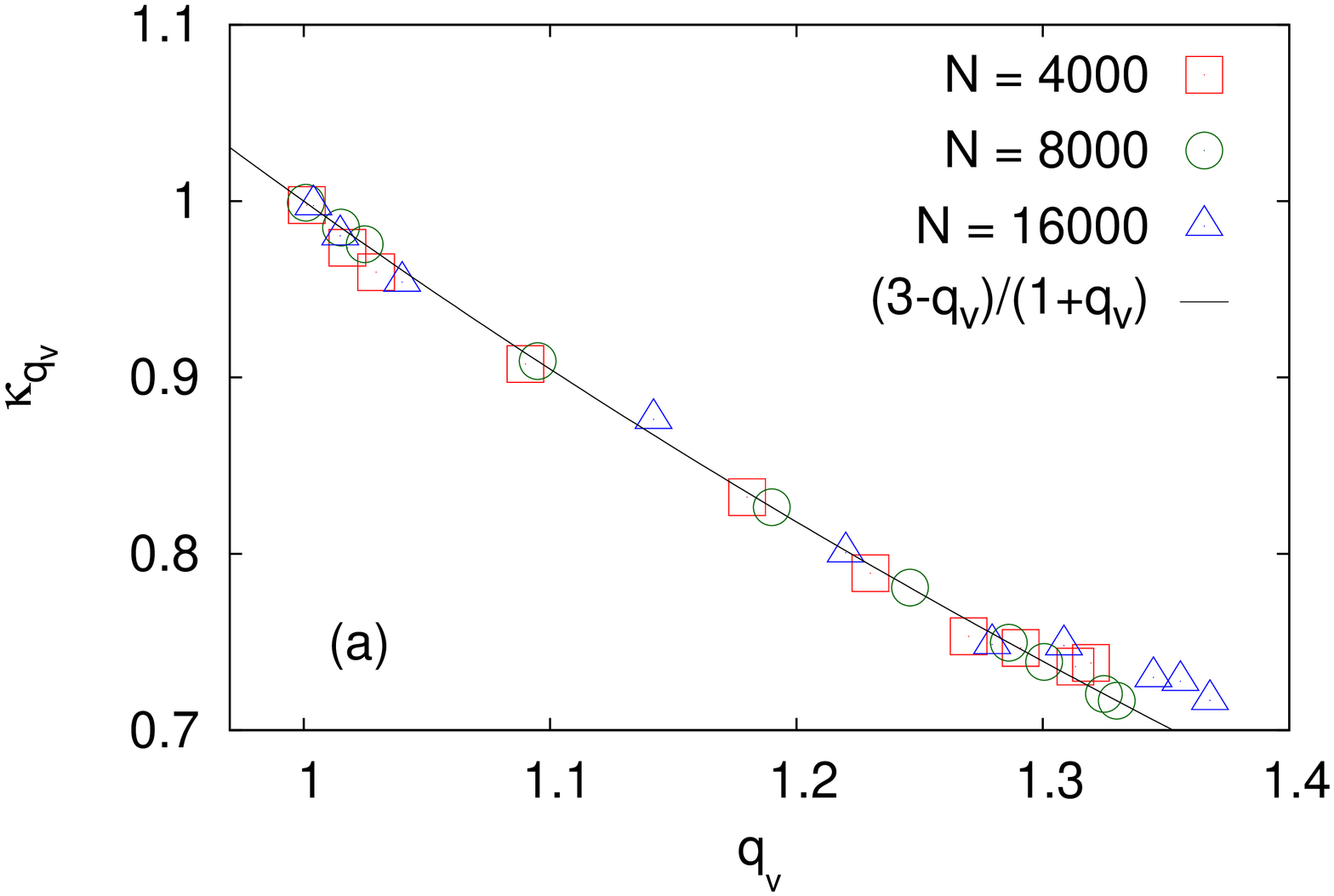}
\includegraphics[width=5cm,angle=-90]{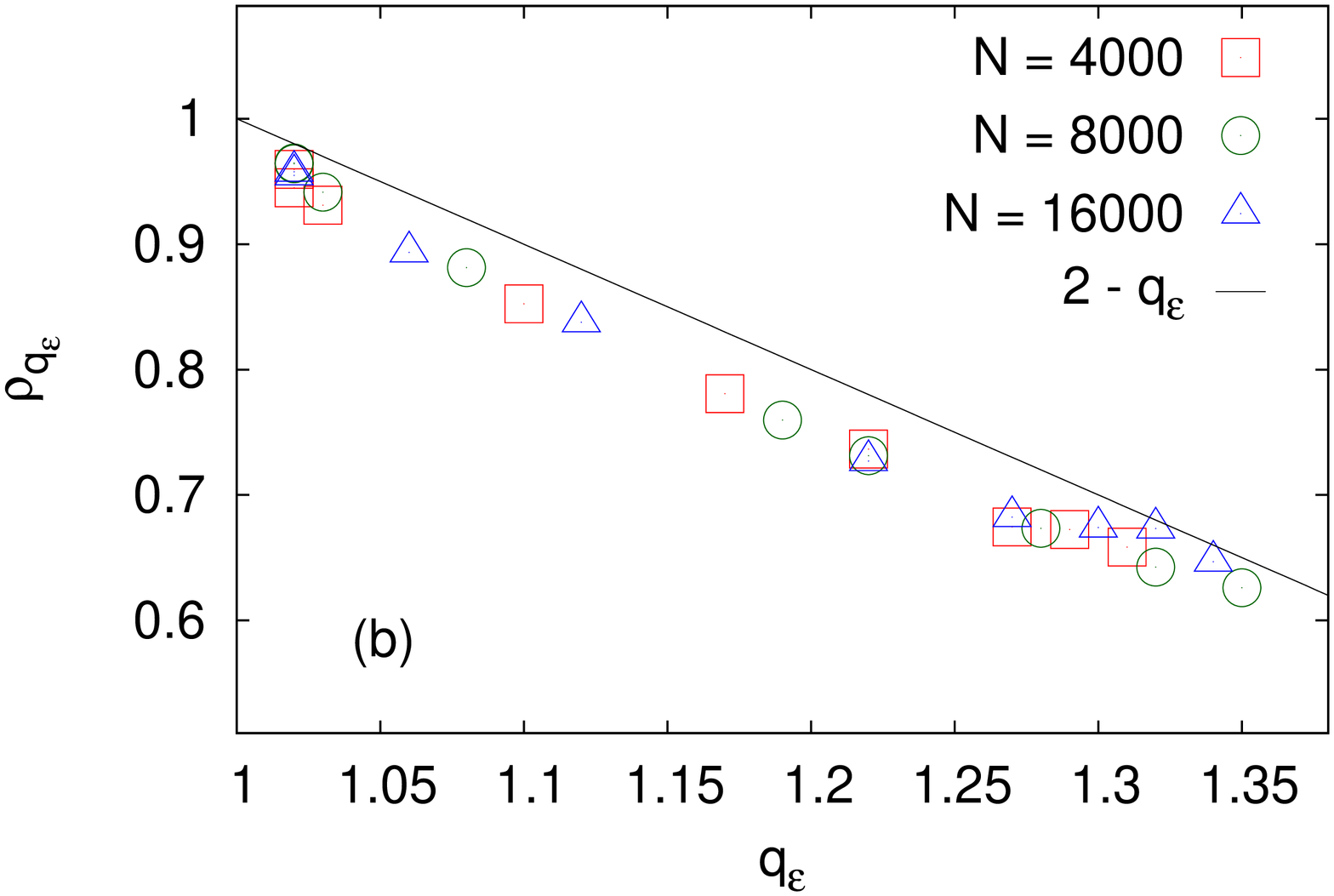}
\caption{(a) $q$-kurtosis for the velocity indices $q_v$: we verify that the computational data satisfactorily follow the 
analytical result $\kappa_{q_v}=(3-q_v)/(1+q_v)$ corresponding to $q_v$-Gaussians. (b) $q$-ratio for the energy indices 
$q_{\epsilon}$:  the computational data are close to the expected relation $\rho_{q_{\epsilon}} = 2-q_{\epsilon}$. These 
two results reinforce the $q_v$-Gaussian and $q_\epsilon$-exponential ansatzes for fitting the velocity and energy histograms. 
The fact that the values of $\rho_{q_\epsilon}$ are slightly lower than their analytical expectation comes from the fact that
we have not taken into account the fact that, at low energies (of the order of unit), there is a departure from the pure 
$q_\epsilon$-exponential behavior due to the density of states, similarly to what generically happens in the Boltzmannian 
regime of any model.  
}
\label{qkurtosis}
\end{figure}

The adequacy of the $q$-Gaussian and  $q$-exponential forms has been checked with quantities such as the $q$-kurtosis $\kappa_q$
and $q$-ratio $\rho_q$ respectively. From the velocity histograms we can compute the $q$-kurtosis $\kappa_q$ of the distribution,
defined as \cite{TsallisPlastinoEstrada2009,CirtoAssisTsallis2013,ChristodoulidiTsallisBountis2014}
\begin{equation}
\kappa_q = \frac{\int_{- \infty}^{\infty} dv ~  v^4 [P(v)]^{2 q -1} / \int_{- \infty}^{\infty} dv [P(v)]^{2 q -1}}{3 \left[ \int_{- \infty}^{\infty} dv ~  v^2 [P(v)]^{q} / \int_{- \infty}^{\infty} dv [P(v)]^{q} \right]^2};
\label{q-kurtosis}
\end{equation}
the  value of $q$ is obtained by fitting the velocity histograms obtained from simulation. Using Eq. (\ref{q-kurtosis}) 
the $q$-kurtosis of any histogram can be computed. In particular, it can be verified that $\kappa_q = (3-q)/(1+q)$
for any $q$-Gaussian velocity distribution
\begin{equation}
P(v) = A_v \left[1 - \beta_v (1-q) v^2/2 \right]^{1/(1-q)} \;\;\;(A_v > 0;\, \beta_v >0) \,.
\label{q-G}
\end{equation}
Note that, for $q\to 1$, we recover the well known kurtosis $\kappa_1 = \langle x^4 \rangle/ 3 \langle x^2 \rangle^2 =1$
mandated by Gaussian distributions.
Analogously, from the energy histograms we can compute the $q$-ratio $\rho_q$ defined as follows
\begin{equation}
\rho_q = \frac{\int_0^{\infty} d\epsilon ~  \epsilon^2 [P(\epsilon)]^{2q-1} / \int_0^{\infty} d\epsilon [P(\epsilon)]^{2q-1}}{2 \left[ \int_0^{\infty} d\epsilon ~  \epsilon [P(\epsilon)]^{q} / \int_0^{\infty} d\epsilon [P(\epsilon)]^{q} \right]^2};
\label{q-ratio}
\end{equation}
the  value of $q$ is obtained by fitting the energy distributions. Using Eq. (\ref{q-ratio}) the $q$-ratio of any histogram 
can be computed. In particular, it can be verified that $\rho_q = 2-q$ for any $q$-exponential energy distribution
\begin{equation}
P(\epsilon) = A_\epsilon \left[1 - \beta_\epsilon (1-q) \epsilon \right]^{1/(1-q)} \;\;\;(A_\epsilon>0;\, \beta_\epsilon >0) \,.
\label{q-E}
\end{equation}
Note that, for $q\to 1$, we recover the well known ratio $\rho_1 = \langle \epsilon^2 \rangle/ 2 \langle \epsilon \rangle^2 =1$ 
mandated by exponential distributions. The numerical data obtained for the $q-$kurtosis and the $q-$ratio are displayed in 
Fig. \ref{qkurtosis},a and b respectively, along with their analytical expressions.

\begin{figure}
%\centering
\includegraphics[width=5cm,angle=-90]{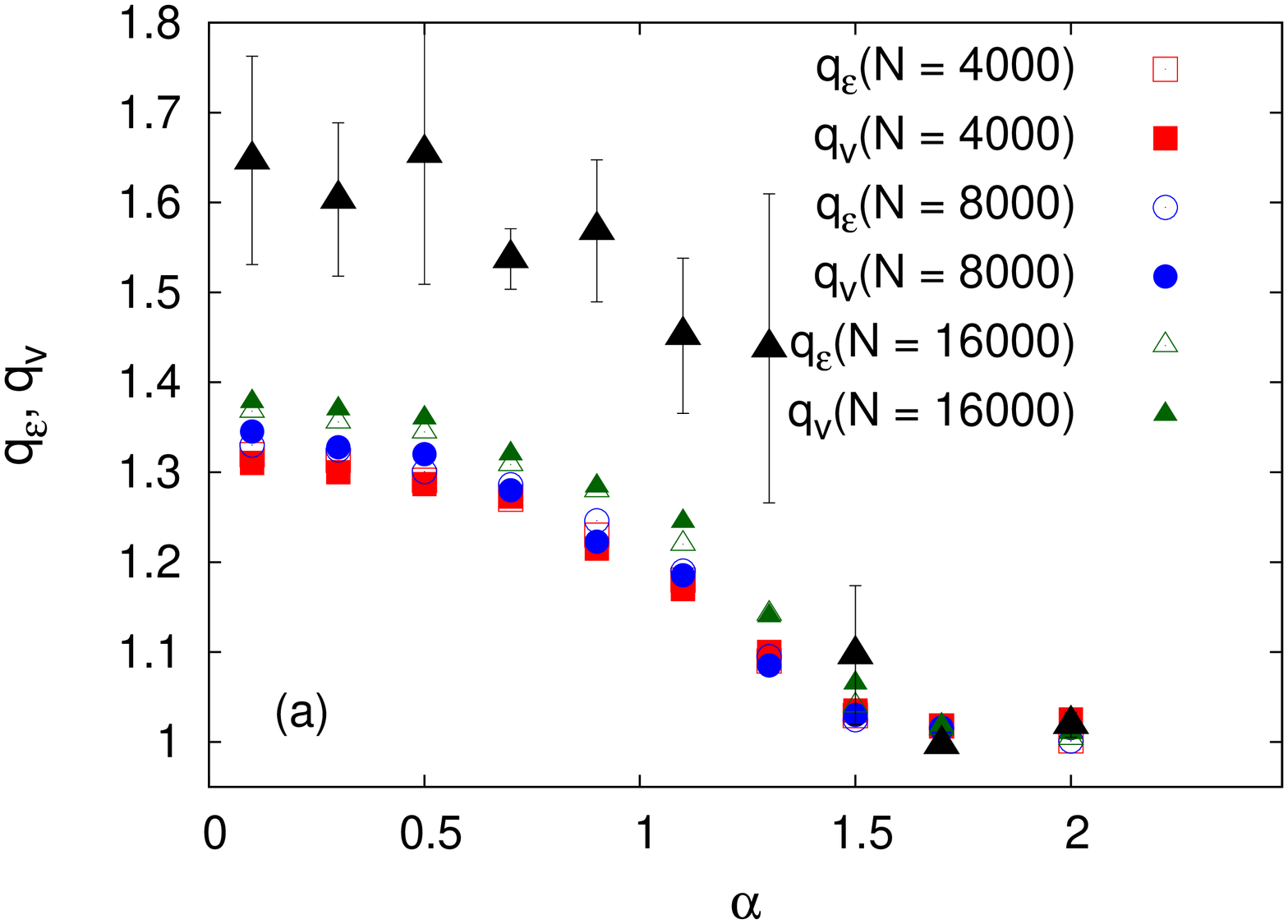}
\includegraphics[width=5cm,angle=-90]{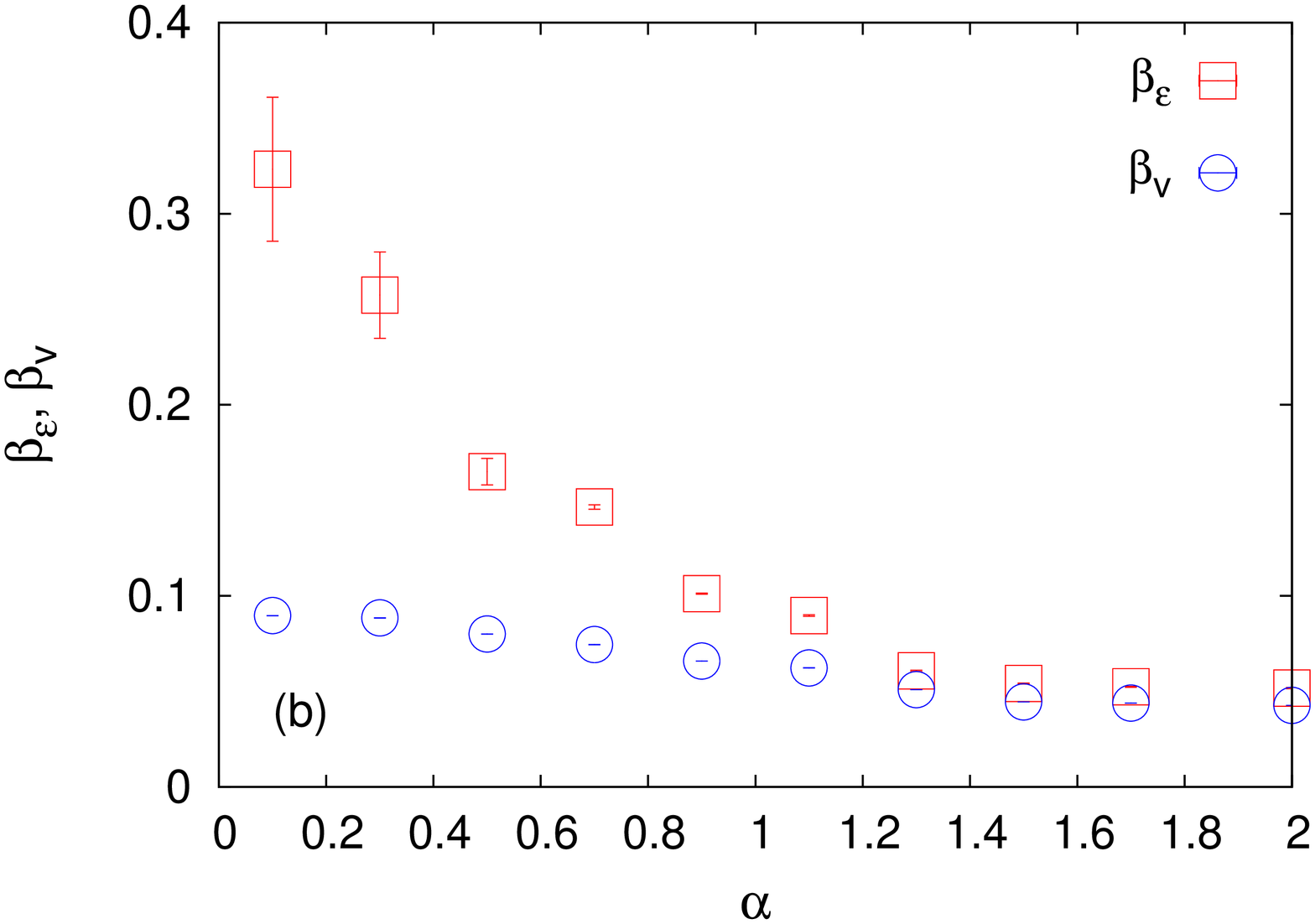}
\caption{(a) We exhibit here that, within error bars, $q_{\epsilon}=q_v$ monotonically decreases for increasing $\alpha$ and 
achieves the Boltzmann value $q_{\epsilon}=q_v=1$ for short-range interactions. The $N \to\infty$ values (black triangles) have 
been extrapolated from the finite $N$ values by following the procedure indicated in \cite{ChristodoulidiBountisTsallisDrossos2016}, 
namely performing $q(N)$ versus $1/\ln N$ extrapolations for increasing $N$.
(b) We exhibit here that, in contrast with the corresponding $q$-indices, the inverse temperatures $\beta_\epsilon$ and 
$\beta_v$ {\it do not} coincide unless we are in the Boltzmannian regime $q_v=q_{\epsilon}=1$ (short-range interactions); 
the emergence of temperatures which differ from the usual kinetic one are frequent in such complex systems
\cite{Andradeetal2010,CirtoAssisTsallis2013}
}
\label{qindices}
\end{figure}
The results for the $q$-indices (as well as for the associated inverse temperatures $\beta$ 's), presented in Fig. \ref{qindices}a
and b are very eloquent, and deserve some comments. The numerical values for the indices $q_{\epsilon}$ and $q_v$ {\it coincide}
within small error bars, for all the values of $(N,\alpha)$ that have been computationally run. This fact surely is nice and simple
if we take into account the fact that both indices are associated with one-variable {\it marginal} distributions coming from the 
same many-body distribution (in a $2N$-dimensional phase space), which is of course numerically inaccessible for the large values 
of $N$ that have been used in the present calculations. Naturally, the precise $q$-exponential and $q$-Gaussian forms fail  in the
numerical regions of too high one-body energies $\epsilon$ and too high velocities $|v|$ respectively. The goal is in fact to attain 
as best as possible the $N\to\infty$ limit, where the analytical forms could possibly be correct for all energies and all velocities.
Such extrapolations have been done as shown in Fig. \ref{qindices}a: they exhibit $q \simeq 5/3$ for the $\alpha=0$ limit, and a 
monotonically decreasing value for $q$ when $\alpha$ increases up to $\alpha_c$, and $q=1$ for $\alpha>\alpha_c$, with $\alpha_c 
\simeq 1.6$. The arguments based on the positivity of the largest Lyapunov exponent (see, for instance, \cite{TirnakliBorges}) rather
suggest $\alpha_c=1$. However, all the related numerical results available up to now in various classical $d$-dimensional models 
\cite{CirtoAssisTsallis2013,ChristodoulidiTsallisBountis2014,ChristodoulidiBountisTsallisDrossos2016,CirtoRodriguezNobreTsallis2017} 
systematically and intriguingly indicate $\alpha_c/d \equiv a_c >1$. This robust numerical fact remains so even if $N$, time, precision, 
integrating algorithms and other circumstances, are modified. This unexpected peculiarity has remained irreducible up to now, and it 
might suggest a distinction between {\it strongly long-range-interacting} systems ($0 \le \alpha/d < 1$) and {\it weakly long-range-interacting} 
systems ($1 < \alpha/d < a_c$), with $a_c$ roughly in the range 
$(1.5,
2)$; the strictly short-range-interacting systems would therefore correspond to $\alpha/d > a_c$. This situation is somehow reminiscent
of say the $d=1$ quantum Ising ferromagnet with long-range interactions, which is known to present {\it three} (and not only {\it two}) 
thermostatistical regimes \cite{ThreeRegimes}, namely $0<\alpha<1$, $1<\alpha<2$, and $\alpha>2$, corresponding therefore to $a_c=2$. 
The physical interpretation of the present most interesting three regimes remains elusive. However, one possibility could be that the 
emergence of a neat Boltzmann regime (i.e. $q=1$) in a microcanonical ensemble demands not only ergodicity over the entire phase space 
(obviously assured by a positive largest Lyapunov exponent), or over a nonvanishing-Lebesgue-measure part of it, but, in addition to that, 
an uniform probability distribution over that region of phase space. If so, the thermostatistical regime for $1< \alpha/d < a_c$ would of 
course ultimately satisfy ergodicity, but not equal probabilities effectively (at least not yet at the largest times that have been 
computationally attained). The deep understanding of this point would further enlighten the first-principle conditions of validity of the 
celebrated Boltzmann-Gibbs statistical mechanics.

\vskip0.5cm
{\bf Acknowledgments:}
We acknowledge fruitful discussions with L.J.L. Cirto, E.M.F. Curado, F.D. Nobre, A.R. Plastino, P. Rapcan, G. Ruiz, G. Sicuro, A.M.C. 
Souza, U. Tirnakli and R. Wedemann, as  well as partial financial support from CNPq and Faperj (Brazilian agencies), and from the John 
Templeton Foundation (USA).

\end{document}